\documentstyle[twoside,fleqn,espcrc2,psfig]{article}

\newcommand{\AmS}{{\protect\the\textfont2
  A\kern-.1667em\lower.5ex\hbox{M}\kern-.125emS}}
\newcommand{\pomeron}{{\protect\rm l\hspace*{-0.15em}P}}

\title{
Intermittency analysis in diffractive DIS processes
}

\author{Zhang Yang\\
Institut f\"ur theoretische Physik, FU Berlin,
 Arnimallee 14, 14195 Berlin, Germany}

\begin{document}

\begin{abstract}
It is pointed out in this talk that ``the colorless objects'' 
in diffractive  scattering in the small-$x_B$
region can be probed by 
measuring the scaled factorial moments of final-state-hadrons and the
dependence of their scaling behavior 
upon the diffractive kinematic variables.
The Monte Carlo implementation of RAPGAP and JETSET are discussed as
illustrative examples. The results of these model-calculations show in
particular that inclusion of the contributions from
the photon gluon fusion processes can considerably enlarge the power of the
scaled factorial moments.
The significant $x_B$-dependence of intermittency index in small-$x_B$
region in this MC analyses shows the hadronic final states depend very
much on the lifetime of the exchanged colorless object $c_0^*$.
The dependence of the
intermittency index upon $x_\pomeron$, $Q^2$ and $\beta$ are also discussed.

\end{abstract}

\maketitle

Recent years have witnessed a remarkably intense experimental and
theoretical activity in research of the dynamics of diffractive
deep-inelastic electron-proton (ep) scattering (DIS) in low
Bjorken-$x$ kinematic range. The measurements at HERA have
demonstrated the existence of a distinct class of events in which
there is no hadronic energy flow in an interval of pseudo-rapidity
adjacent to the proton beam direction\cite{m1}. Our present
understanding of DIS could be inadequate at low-$x_B$ because additions to
the leading order QCD-based partonic picture are likely to be
substantial. A natural interpretation of these so-called ``large rapidity
gap'' events is based on the hypothesis that the deep-inelastic
scattering process involves the interaction of the virtual boson probe
with a colorless component of the proton. Hence there is no
chromodynamic radiation in final state immediately adjacent to the
direction of the scattered proton or any proton remnant. What is 
this colorless component originating from the nucleon? 
The large rapidity gap events discovered in deep inelastic scattering at
HERA\cite{m1} are usually interpreted in terms of pomeron-exchanged
model\cite{m2}. Although this seems to work reasonably well
phenomenologically, there is yet no satisfactory understanding of the
pomeron structure and its interactions mechanism. 
In this respect, it is helpful to
probe the properties of such objects in the non-traditional aspects, in
addition to the traditional measures such as rapidity gap
distribution, structure function and the 
averaged cross section and so on\cite{m1}.
In the present talk we wish to point out that useful information about the
exchanged colorless 
object in diffractive lepton-nucleon scattering process can
be extracted by studying scaling behavior and fractality
(intermittency) of the final
state of the colorless component in proton excited by the virtual
boson probe.

The manifestation of
fractality (intermittency) in high energy
multiparticle production process is the anomalous scaling\cite{bialas,kittel}
\begin{equation}
\label{e1}
F_q(\delta x)=F_q(\Delta x)\left({\Delta x\over \delta x}\right)^{\phi _q},
        \  {\rm as}\ M\to \infty,\delta x\to 0
\end{equation}
of $q$-order factorial moments (FM's) $F_q$, defined as
\begin{equation}
\label{e2}
F_q(\delta x) = {1\over M}\sum ^M _{m=1}
    {\langle n_m(n_m-1)\dots (n_m-q+1)\rangle \over
    \langle n_m \rangle ^q},
\end{equation}
where $x$ is some phase space variable, e.g. (pseudo-)rapidity, 
the scale $\delta x=\Delta x/M$
is the bin width for a $M$-partition of the region $\Delta x$ in
consideration, $n_m$ is the multiplicity in the $m$th bin.
Since the factorial moments can remove the statistic noise around
the probability and associated directly with
the scaled moments of probabilities\cite{bialas},
the scaling exponent $\phi_q$ in Eq.(~\ref{e1}), called intermittency 
index, characterizing the strength of dynamical fluctuation, is
connected\cite{lipa} with the anomalous fractal dimension $d_q$ of
rank $q$ of the spatial-temporal evolution of high energy collisions, 
$d_q={\phi _q/(q-1)}$. A possible
cause leading to the power-law of FM's of final particles 
in high-energy collisions is that the
emitting source of final hadrons is self-similar fractal\cite{bialas1}.
Furthermore, DIS experiments and the empirical analyses\cite{m1,m3}
 show that the
gluon-density  in the nucleon  in the low-$x_B$ kinematical region is 
much higher than those for quarks/antiquarks, and it is 
increasing with decreasing $x_B$\cite{m3}.
In this soft gluon system where the gluons interact with each other in
complicated dynamic processes, it has been
proposed\cite{soc1} that the gluons may 'self-organize' into the
clusters in the dissipative gluon system by
self-organized criticality (SOC)\cite{soc2}, 
and the colorless component in proton can be regarded as color
singlet gluon clusters\cite{m4}. The spatial-temporal
structure of the SOC-cluster (or BTW-cluster) is self-similar
fractal\cite{soc2}. So it is
feasible to study the fractal structure of the colorless component of
the proton by measuring the anomalous scaling behaviors of factorial
moments of the final state particles originating from the scattering 
of virtual photon and colorless object in the
diffractive lepton-nucleon scattering.

The $x_B$-dependence of scaling behaviors of FM's, and in particular
that of intermittency index 
$\phi_q(Q^2, x_B; x_\pomeron)$ for fixed $Q^2$ (and $x_\pomeron$), 
plays a distinguished role in such studies. This is because,
when the virtual photon $\gamma^*$ originating from the incident lepton is
absorbed by the nucleon,
its energy-momentum $q\equiv (q^0,
\vec q)$ is in fact absorbed by a virtual colorless component of the
nucleon in the lepton-nucleon diffractive scattering.
In a fast moving reference frame, for example the lepton-nucleon
center-of-mass frame, where the nucleon's momentum $\vec P$ is large 
in high-energy collisions,
the time interval $\tau_{\rm int}$ in which the absorption
process takes place (it is known as the lepton-nucleon 
interaction/collision time)
can be estimated by making use of
the uncertainty principle. In fact, by calculating 
$1/q^0$ in this reference
frame we obtain\cite{formula}:
\begin{equation}
\label{e3}
\tau_{\rm int} \sim {4|\vec P| \over Q^2} ~ {x_B\over 1-x_B}.
\end{equation}
This means, for given $\vec P$ and $Q^2\equiv -q^2$, 
the interaction time $\tau_{\rm int}$ is directly proportional to
$x_B$  in the low-$x_B$ ($x_B\ll 1$) kinematic range. 
In other words, $x_B$ is a measure of
the time-interval in which the absorption of $\gamma^*$
by the space-like virtual colorless
object takes place.
Hence, by studying the $x_B$-dependence of the intermittency
index $\phi_q(Q^2,x_B;x_\pomeron)$, we  are 
not merely probing the anomalous scaling behaviors
of the collision process between $\gamma^*$ and 
the colorless object which we hereafter call $c_0^*$.
Since this hadronization process of the virtual colorless object $c_0^*$
is initiated by
the interaction with $\gamma^*$,
we are also examing
whether/how the hadronization process 
changes with the interaction time $\tau_{\rm int}$. 
This question is of considerable interest, because 
a virtual photon $\gamma^*$
can (logically) only be absorbed by virtual systems ($c_0^*$'s) whose
lifetimes ($\tau_c^*$'s) are longer than interactive time 
$\tau_{\rm int}$ (i.e. $\tau_{\rm int}\le\tau_c$). 
That is, the average lifetime $\langle
\tau_c\rangle$ of the $c_0^*$'s, which can absorb a $\gamma^*$
associated with interaction-time $\tau_{\rm int}$,
is a function of $\tau_{\rm int}$.
Hence, from the $x_B$-dependence of the scaling behavior of FM's,
in particular from that of the corresponding $\phi_q(x_B, Q^2;
x_\pomeron)$'s, 
we can in principle find out whether/how
the dynamic fluctuation and the fractal structure in the
spatial-temporal evolution originating from
the exchanged colorless object $c_0^*$ depends on its average
lifetime $\langle\tau_c\rangle$ of the $c_0^*$'s.

The $Q^2$-dependence of
the scaling behavior of the FM's is also of considerable interest.
This is because, in photon-proton scattering experiments, 
not only those with real $(Q^2=0)$ photons but also
those with space-like $(Q^2>0)$ photons where $Q^2$ is not 
too large ($\le 1 {\rm GeV}^2/c^2$, say)
have very much in common with hadron-hadron collisions.
Having in mind that the index of intermittency
for hadron-hadron scattering is smaller than that for 
electron-positron annihilation processes\cite{kittel},
we are led to the following questions: Do we expect
to see a stronger $Q^2$-dependence when we increase $Q^2$ from zero to 
10 or 100 $GeV^2/c^2$, say? Is this also a way to see whether space-like
photons at large $Q^2$ "behave like hadrons" in such interactions?

While waiting for data to perform the above-mentioned analysis, 
let us now consider the following two phenomenological models 
as illustrative examples:

(A). If the colorless object ($c_0^*$) is a
quark-antiquark pair (formed by interacting 
gluons) which exists 
in the time-interval when the virtual photon $\gamma^*$
is absorbed by the object, we shall see the following: Especially when
$Q^2$ is sufficiently large, the incoming $\gamma^*$ (the transverse dimension
is expected to be proportional to
$1/Q^2$) will hit the quark $(q)$ or the antiquark $(\bar q)$ and make them
fly apart symmetrically
 with respect to the center of mass
 the $\gamma^*q\bar q$ system --- similar to the
$q\bar q$-pair produced in $e^+e^-$-collisions (with respect to  the 
center of mass the $q\bar q$ 
system). That is, 
in this case, the final-state-hadrons of an event are 
fragmentation-products of the quark and/or the
antiquark, and hence they are
expected
to show characteristic features similar as 
those observed in the reaction 
$e^+e^-\to $ hadrons at the
same c.m.s. energy. It is interesting
to see that the very recent inclusive measurements performed 
at HERA\cite{m1} in diffractive electron-proton scattering show 
the following: Both the scaled longitudinal momentum ($x_F$) distribution
and the transverse momentum ($p_\bot$) distribution are strikingly symmetric
with respect to the center-of-mass of the photon and the struck colorless 
object; and the general features
of these distributions are very much the same
as those observed in electron-position collision processes. These facts 
strongly suggest that a more detailed comparison between
these two collision processes would be useful.

For this purpose,
we made use of the Monte Carlo (MC) program JETSET\cite{jetset}. We
generated 50,000 MC
events, and calculated the second normalized factorial
moment in 3-dimensional 
($\eta,p_\bot,\phi )$ phase space at the given $q\bar q$-c.m.s
energy $\sqrt s$, where the pseudorapidity $\eta$, transverse momentum
$p_\bot$ and the azimuthal angle $\phi$ are defined with respect to
the sphericity axis of the event. The usual cumulative variables $X$
translated from $x=(\eta,p_\bot,\phi )$, i.e.
$X(x)=\int^x_{x_{\rm min}}\rho (x)dx/\int^{x_{\rm max}}_{x_{\rm min}}
\rho (x)dx$,
were used to rule out the enhancement of $F_q$ from 
a non-uniform inclusive spectrum $\rho (x)$ of the final hadrons\cite{z3}.
The obtained results have been collected in Fig.1a in
double logarithmic $F_2$
vs M plots for different $\sqrt s$ (or $M_X$ , which 
is the invariant mass of the $\gamma^*
c_0^*$ system in a corresponding 
diffractive lepton-nucleon scattering event).
It is clear that the higher the invariant mass of $\gamma^*c_0^*$
is, the larger the power of FM's, i.e. 
the stronger the dynamic fluctuation is. In the very low invariant mass
($\sqrt s=4.5$GeV, say), the powers of FM's become less than 0, which
can be referred to the constraint of the momentum conservation in the
hadronization process\cite{zhang}.

\vskip 0.5cm
\begin{figure}[vhb]
\vskip -1.5cm
\centerline{
\psfig{figure=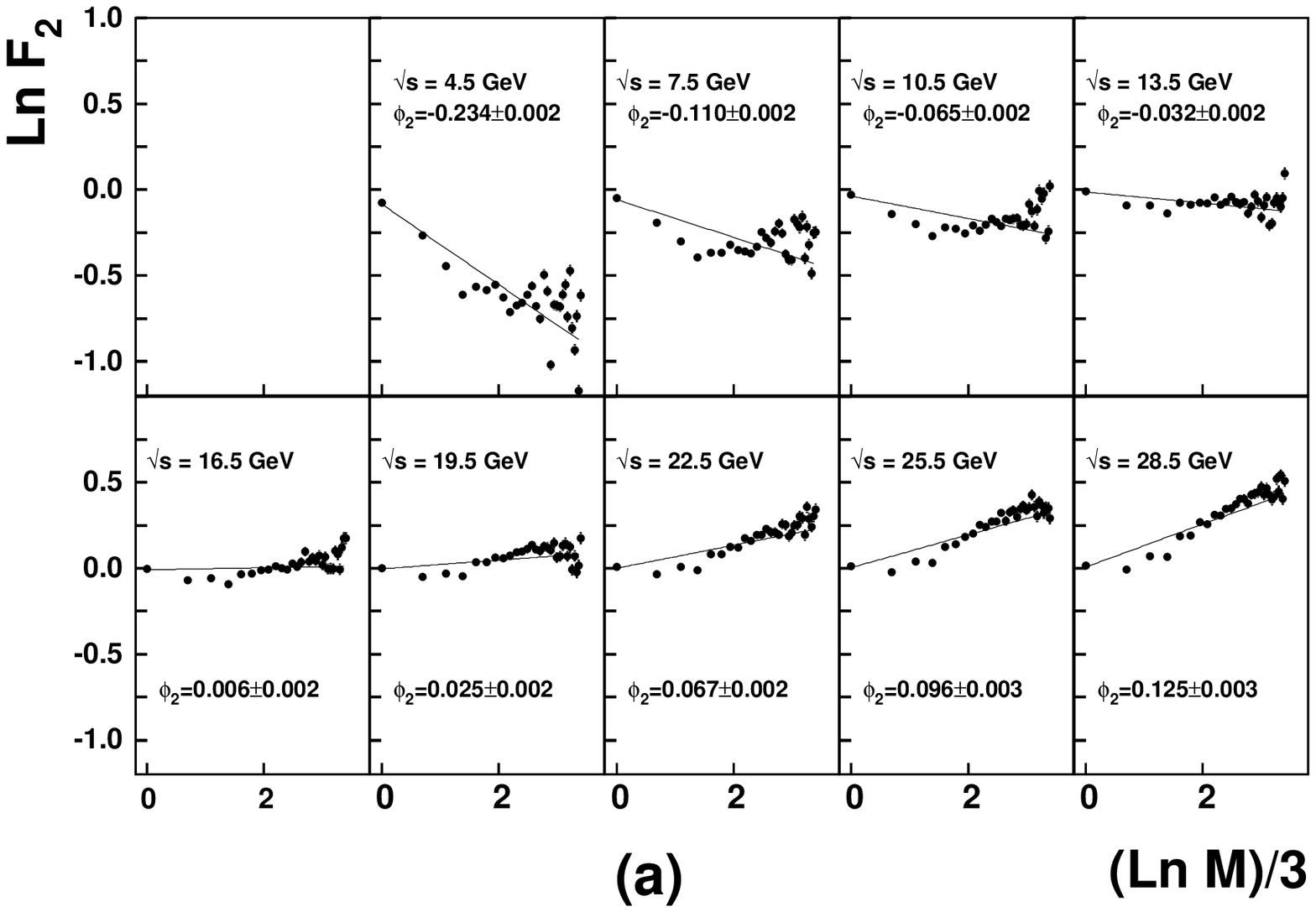,width=9cm}
}
\vskip -2.5cm
\end{figure}
\mbox{}
\begin{figure}[vhb]
\vskip -1.5cm
\centerline{
\psfig{figure=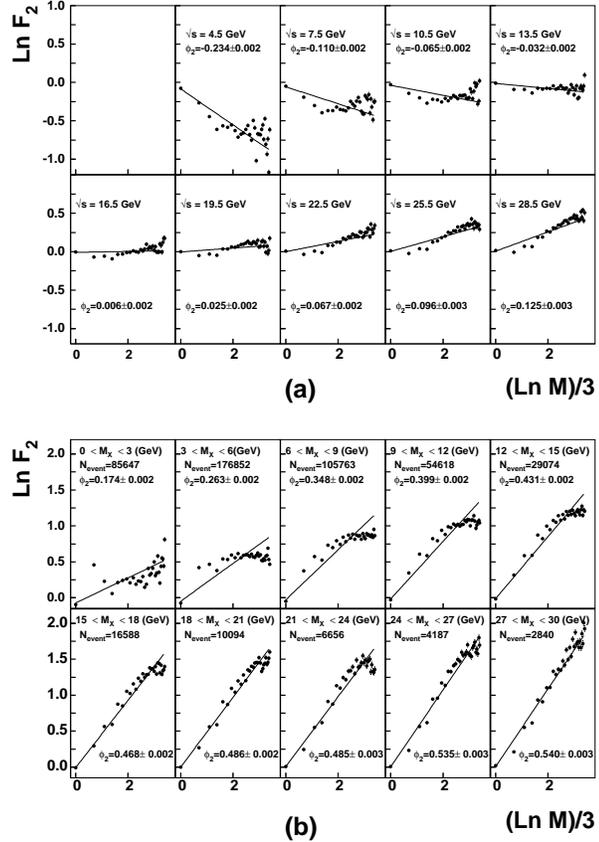,width=9cm}
}
\vskip -2.2cm
\caption{
The scaled factorial moments $F_2$ versus the number $M$ of
subintervals of 3-dimensional ($\eta,p_\bot,\phi$) phase space in
log-log plot, 
and the second-order intermittency index $\phi_2$ in corresponding
sample set.
(a). The MC result of JETSET 7.4{\protect\cite{jetset}} in different
cms energy ${{\protect\sqrt s}}$, 
50000 events are generated in each sample set;
(b). That of RAPGAP{\protect\cite{rapgap}} in corresponding invariable
mass $M_X$ interval with $N_{\rm event}$ events in each subsample.
}
\end{figure}

\vskip -0.5cm

\mbox{}
(B). \mbox{} Based on pomeron-exchanged model\cite{m2}, 
several \  Monte\  Carlo\  generators,\  such  \ as POMPYT\cite{pompyt}, 
RAPGAP\cite{rapgap} have been set up. 
These models have been used to describe 
quite well the HERA data in the global and averaged features, such as
the rapidity gap distribution, the
diffraction and total cross section and so on\cite{m1}.
The theoretic models are usually confronted with incorrigible discrepancy,
when the data about the locally nonstatistic fluctuation in small
phase space are involved in the comparison with them\cite{kittel}. 
No evidence has shown that this kind of
locally dynamic fluctuation could be certainly referred only to the
hadronization process but have nothing to do with the initial stage of
high energy collisions. 
In this respect, it is
also relevant to see whether the features of scale invariance and
fractality in small phase space of the lepton-nucleon diffractive
processes, specially their dependence upon the diffractive
variables can be compatible with the pomeron type of model. 
In the following we take RAPGAP as an example, in which the
virtual photon ($\gamma^*$) interacts directly with a parton
constituent of the pomeron either in lowest order (Fig.2a) or in
$O(\alpha_{\rm em}\alpha_s)$ order --- via the photon gluon 
fusion (Fig.2b). In both cases a color octet remnant is left at low
$p_T$ with respect to the pomeron and hence also to the proton. The
higher order gluon emission is simulated with the Color Dipole Model
(ARIADNE)\cite{ariadne} and the hadronisation is performed using the
JETSET\cite{jetset}. The pomeron flux factor
$f_{P\pomeron}(t,x_\pomeron)$ and  
pomeron structure function $G(\beta )$ are taken respectively as
\begin{equation}
\label{e4}
f_{P\pomeron }(t,x_\pomeron)={\beta^2_{P\pomeron }(t)
\over 16\pi}\beta^{1-2\alpha_\pomeron (t)},
\end{equation}
given by Berger et al. and Streng\cite{m2}, and
\begin{equation}
\label{e5}
\beta G(\beta)=6 \beta (1-\beta),
\end{equation}
when the pomeron is made of 2 `unrealistically hard' gluons as suggested by
 Ingelman and Schlein\cite{m2}.
Here the kinematic variables $t$, $x_\pomeron$ and $\beta$ are defined
as $t=(P-P')^2$, $x_\pomeron={q\cdot (P-P')/ (q\cdot P)}$, 
and $\beta ={-q^2/[2q\cdot (P-P')]}$, where $P'$ is the 4-momentum
of the final state colorless remnant of nucleon.
The pomeron Regge trajectory is given by
$\alpha_\pomeron (t)=1+\epsilon +\alpha 't$, $\epsilon\approx
0.085$ and the slope $\alpha '=0.25$ obtained from a fit to
data\cite{m2}. 

\newpage
\mbox{}
\begin{figure}
\hskip -0.5cm
\centerline{
\psfig{figure=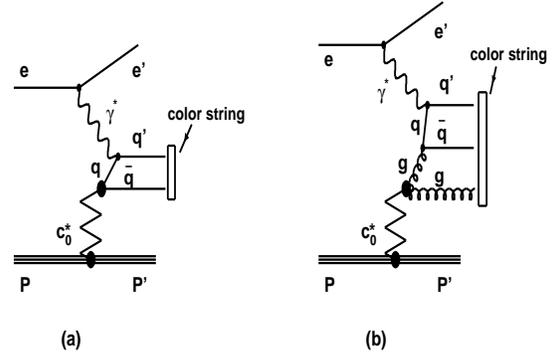,height=8cm,width=9cm}
}
\vskip -2.2cm
\caption{ 
The basic processes included in the RAPGAP{\protect\cite{rapgap}} 
implementation for 
inelastic lepton scattering on a pomeron: (a) the lowest order process
for hard parton level; (b) the $O(\alpha_{\rm em}\alpha_s)$ order
process for photon gluon fusion.
} \label{fig2} \end{figure}


\mbox{}
\vskip -.5cm

We generated 500,000
RAPGAP events, and divided the whole sample into 10 subsamples according to
the invariable mass $M_X$ of $\gamma^*c_0^*$ in the MC event. 
The scaling behaviors of FM's for different $M_X$ intervals are shown in
Fig. 1b. The dependence of scaling behaviors of FM's upon $M_X$ is
similar with the  result of JETSET in Fig.1a, i.e. the powers increase
for increasing scattering energy of $\gamma^*c_0^*$.
But it is noticeable that the intermittency index $\phi_2$ for a given
$M_X$ interval is much larger in RAPGAP than that in JETSET.
Having in mind that in Fig. 1a
the colorless object is considered as a quark-antiquark pair
formed by gluons and the Feymann graph of the diffractive process in
this aspect is just
same as shown in Fig. 2a, the difference between cases (A) and (B) is
that the higher-order photon-parton interaction
has been taken into account in RAPGAP (see Fig. 2b). 
It is understandable since the branch
number of the parton cascading process in parton shower level is
larger when the photon gluon fusion is involved in Fig. 2b, 
while it is believed generally that\cite{kittel} power-law
behaviors in the 
color-string type of models are referred in large part to the
randomly cascading process of parton energy in perturbative phase,
so the fractal in the evolution processes with longer cascade branch 
would be stronger.

\mbox{}
\bigskip

\begin{figure}
\centerline{
\psfig{figure=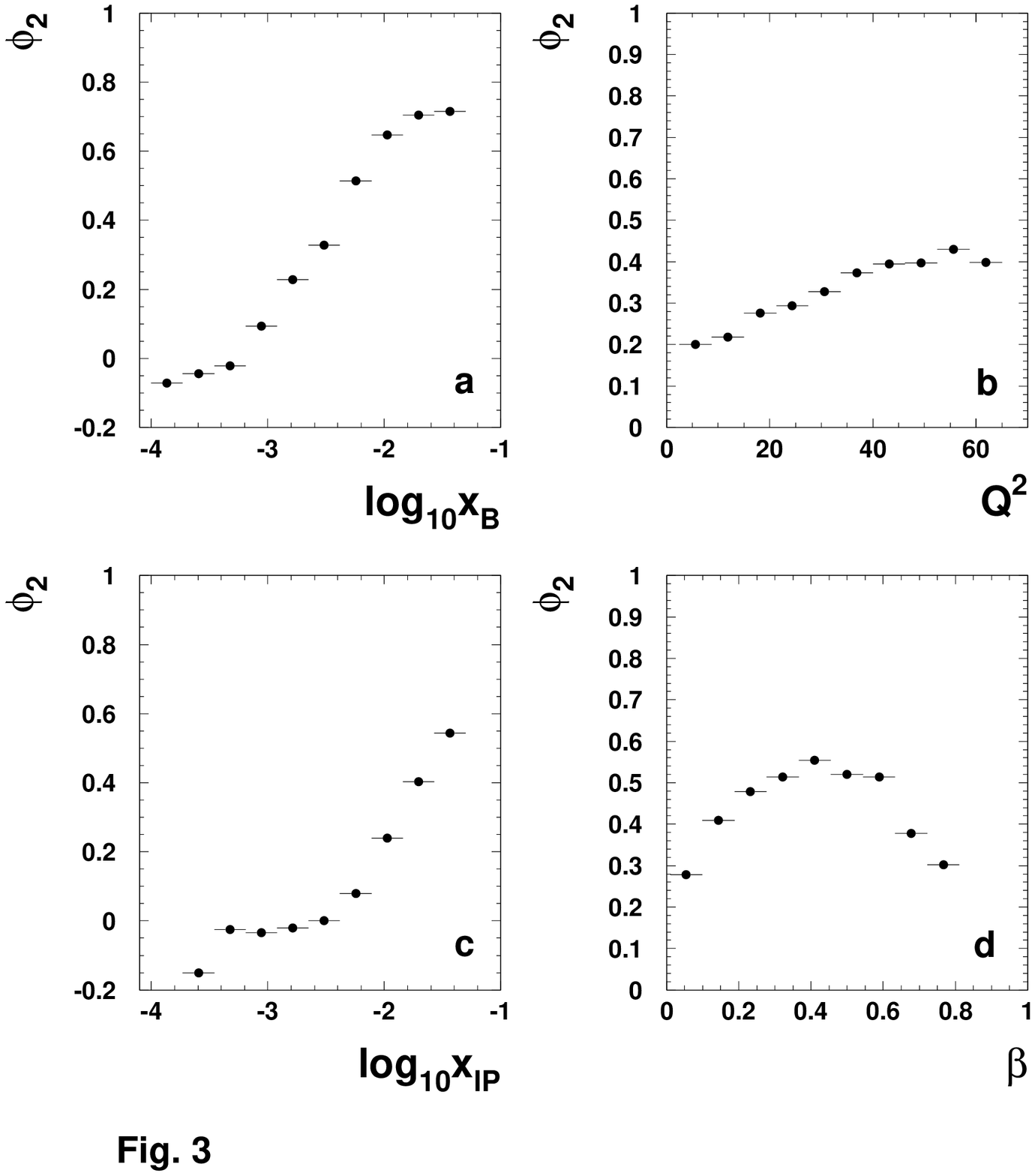,height=11cm,width=9cm}
}
\vskip -2.2cm
\caption{ 
The dependence of 2-order intermittency index $\phi_2$ in
RAPGAP{\protect\cite{rapgap}} 
Monte Carlo implementation upon diffractive kinematic variables: (a) $x_B$, (b)
$Q^2$, (c) $x_\pomeron$ and (d) $\beta$. 
} \label{fig3} \end{figure}


\mbox{}
\vskip -1.2cm

In order to see the dependence of dynamic fluctuation upon the other
diffractive variables as we argued above,  
we divided the RAPGAP MC sample into 10
subsamples according to $x_B$ and $Q^2$.
The results of scaling behaviors of FM's
are shown in Figs. 3a and b respectively.
This example explicitly shows how the proposed method can be 
used to analyze the feature of the evolution precess of the colorless
object in 
diffractive lepton-nucleon scattering processes. The significant 
$x_B$-dependence of the intermittency index observed in this
example not only shows that the hadronic final states
depend very much on the average
lifetime of the exchanged colorless object $c_0^*$ 
if $c_0^*$ can be indeed considered as a pomeron as simulated in
RAPGAP, but also the following:
The longer the average lifetime of $c_0^*$ is, the stronger the
fractal structure of the space-time evolution processes originating
from the $\gamma^*c_0^*$'s interaction will be.
The smaller the $Q^2$ becomes, i.e. the larger transverse dimension of
virtual $\gamma^*$ is, the more $\gamma^*$ ``behaves like a hadron''.

Since $x_\pomeron$ may be interpreted as the fraction of the
4-momentum of the proton carried by the exchanged colorless objects,
and $\beta$ as the fraction of the 4-momenta of the exchanged objects
carried by the parton interacting with the virtual boson, it
meaningful to calculate the dependence of scaling behavior upon
$x_\pomeron$ and $\beta$ (Fig. 3c and d respectively). It seems from
RAPGAP that only when the momentum of colorless object is large
enough, can scaling behavior of the final state
originating from $\gamma^*c_0^*$ be significant;
and it is also clear from Fig. 3d that 
the dynamic fluctuation in the diffractive scattering
don't increase monotonously for increasing the fraction of energy of
stricken parton in the pomeron.

In conclusion, we have shown in this note that the colorless
object $c_0^*$ in diffractive lepton-nucleon scattering can be
probed in a model-independent way, i.e. 
by performing the scaled factorial moment analyses
for the hadronic final states originating from the scattering of
virtual boson $\gamma^*$ and the $c_0^*$ on an event-by-event bases.
The Monte Carlo implementation of RAPGAP and JETSET have been discussed as
illustrative examples. The results of these model-calculations show in
particular that inclusion of the contributions from
the photon gluon fusion processes can considerably enlarge the power of the
scaled factorial moments.
The significant $x_B$-dependence of intermittency index in small-$x_B$
region in this MC analyses shows the hadronic final states depend very
much on the lifetime of the exchanged colorless object $c_0^*$,
i.e. the longer the lifetime of $c_0^*$ is, the stronger the fractal
structure of the space-time evolution processes originating from the
$\gamma^*c_0^*$'s interaction will be.

\newpage
\bigskip
{\Large \bf{Acknowledgments} \rm}
\medskip

Author thanks  C. Boros, M. Derrick, E. De Wolf, D. Gross, Z. Liang,
T. Meng, R. Rittel, K. Schotte and K. Tabelow for helpful discussions,
H. Jung for the patient helps in RAPGAP Monte Carlo generator.
He also wants to thank Alexander von Humboldt Stiftung for the
financial support.


\begin{thebibliography}{99}

\bibitem{m1}
M. Derrick {\it et al.} ZEUS Collaboration, 
Phys. Lett. {\bf B315}, 481(1993);             
Z. Phys. {\bf C70}, 391(1996);                 
T. Ahmed {\it et al.} H1 Collaboration, 
Nucl. Phys. {\bf B429}, 477(1994);             
Phys. Lett. {\bf B348}, 681(1995);             
For a recent review, see e.g. 
N. Cartiglia, {\it Diffraction at HERA}, hep-ph/9703245;
and the papers cited therein.

\bibitem{m2}
G. Ingelman and P. Schlein, Phys. Lett. {\bf B152}, 256(1985);
E. Berger, J. Collins, D. Soper and G. Sterman, Nucl. Phys. 
{\bf B286}, 704(1987);
A. Donnachie and P. Landshoff, Phys. Lett. {\bf B191}, 309(1987);
K. H. Streng, in: Proc. of the Workshop Physics at HERA, Hamburg
(1987), ed. R. D. Peccei;
and the papers cited therein.

\bibitem{bialas}
  A. Bialas and K. Peschanski, Nucl. Phys. {\bf B273}(1986)703;
      {\bf B308}(1988)857.

\bibitem{kittel}
For a recent review, see e.g.,
N. Schmitz, in: Proc. XXI Int. Symp. on Multiparticle Dynamics, Wuhan,
1991, eds. Y. Wu and L. Liu (World Scientific, Singapore, 1992) P.377;
E.A. De Wolf, I.M. Dremin and W. Kittel,
Phys. Rep. {\bf 270}, 1(1996).

\bibitem{lipa}
P. Lipa and B. Buschbeck, Phys. Lett. {\bf B223}(1989)465;
R. C. Hwa, Phys. Rev. {\bf D41}(1990)1456;
Zhang Yang, Liu Lianshou and Wu Yuanfang, Z. Phys, {\bf C71}(1996)499.

\bibitem{bialas1}
A. Bialas, Nucl. Phys. {\bf A545}(1992)285c; 
Acta Phys. Pol. {\bf B23}(1992)561.

\bibitem{m3}
See e.g. 
M. Derrick {\it et al.} (ZEUS Collaboration), 
Phys. Lett. {\bf B345}, 576(1995) and
S. Aid {\it et al.} (H1 Collaboration), Phys. Lett. {\bf B354},
494(1995).

\bibitem{soc1}
C. Boros, T. Meng, R. Rittel and Y. Zhang, `{\it Self-organized
criticality and interacting soft gluons in deep-inelastic scattering}',
preprint hep-ph/9704285.

\bibitem{soc2}
P. Bak, C. Tang and K. Wiesenfeld, Phys. Rev. Lett. {\bf 59},
381(1987); Phys. Rev. {\bf A38}, 364(1988). For a recent review, see
P. Bak, {\it How nature works}(Springer-Verlag, New York, 1996).

\bibitem{m4}
F. E. Low, Phys. Rev. {\bf D12}, 163(1975); S. Nussinov, Phys. Rev. Lett.
{\bf 34}, 1286(1975) and Phys. Rev. {\bf D14}, 246(1976);
C. Boros, Liang Zuo-tang and Meng Ta-chung, Phys. Rev. {\bf D54}, 6658(1996).

\bibitem{formula}
The lifetime $\tau_{\rm int}$ of virtual photon in the lepton-nucleon
center-of-mass frame in DIS can be easily formed in the following simple
kinematic exercise:
$k^{'2}\equiv m_e^2=(k-q)^2=m_e^2+q^2-2k\cdot q$,
where $k\equiv (k^0,\vec k)$, $k'\equiv (k'^0,\vec k')$, $q\equiv
(q^0,\vec q)$  and $P\equiv
(P^0, \vec P)$ are the 4-momentum of incident lepton, outgoing lepton,
exchanged virtual photon and incident nucleon in lepton-nucleon
center-of-mass frame respectively, 
$m_e$ and $m_p$ are the invariance mass of lepton and nucleon. 
Taking into account $\vec k=-\vec P$ and $|\vec k|\gg m_e, |\vec P|\gg
m_p$, then
$m_e^2\simeq m_e^2+q^2-2|\vec P|q^0-2\vec P\cdot\vec q=m_e^2+q^2-2|\vec
P|q^0-2q^0\sqrt{|\vec P|^2+m_p^2}+2P\cdot q$, i.e.
$q^0={2P\cdot q+q^2\over 2\sqrt{|\vec P|^2+m_p^2}+2|\vec P|}\simeq
{2P\cdot q -Q^2\over 4|\vec P|}.$
So the lifetime of virtual photon is 
$\tau_{\rm int}\sim {1\over q^0}\sim {x_B\over 1-x_B}{4|\vec P|\over Q^2}.$

\bibitem{z3}
A. Bialas and M. Gazdzicki, Phys. Lett. {\bf B252}, 483(1990);
W. Ochs, Z. Phys. {\bf C50}, 339(1991).

\bibitem{jetset}
T. Sjostrand, Comput. Phys. Commun. {\bf 39}, 347(1986).

\bibitem{zhang}
Liu Lianshou, Zhang Yang and Deng Yue, Z. Phys. {\bf C73}(1997)535.

\bibitem{pompyt}
P. Bruni and G. Ingelman, DESY-93-187, Proc. Rurophysics Conf. on HEP,
Marseilles 1993, Eds. J. Carr and M. Perottet, P.595.

\bibitem{rapgap}
H. Jung, Comput. Phys. Commun. {\bf 86}(1995)147.

\bibitem{ariadne}
L. L\"onnblad, Comput. Phys. Commun. {\bf 71}(1992)15.


\end{thebibliography}
\end{document}